\documentclass{PoS}

\usepackage{amsmath}
\usepackage{wrapfig}

\title{Charge screening effect on hadron-quark mixed phase in compact stars }

\ShortTitle{Charge screening effect on hadron-quark mixed phase in compact stars }

\author{\speaker{Tomoki Endo} \\ 
        Department of Physics, Kyoto University, Kyoto 606-8502, Japan\\
        E-mail: \email{endo@ruby.scphys.kyoto-u.ac.jp}}

\author{Toshiki Maruyama\\
Japan Atomic Energy Research Institute,\footnote{Present name: Japan
Atomic Energy Agency}\ \ Tokai, Ibaraki 319-1195, Japan \\
	E-mail: \email{maruyama.toshiki@jaea.go.jp}}

\author{Satoshi Chiba\\
        Japan Atomic Energy Research Institute,$^\dagger$ Tokai, Ibaraki 319-1195, Japan\\
	E-mail: \email{chiba.satoshi@jaea.go.jp}}

\author{Toshitaka Tatsumi\\
        Department of Physics, Kyoto University,  Kyoto 606-8502, Japan\\
        E-mail: \email{tatsumi@ruby.scphys.kyoto-u.ac.jp}}

\abstract{We study the charge screening effect in the hadron-quark mixed
phase. By including the charge screening effect, rearrangement of charged
particles occurs and some part becomes locally charge-neutral.
As a result the equation
of state for the mixed phase becomes close to that
given by the Maxwell construction, which means that the
Maxwell construction would effectively gain the physical meaning again
even in the system 
with two or more chemical potentials. We also discuss the interplay of
the surface tension and the Coulomb interaction. Both effects would restrict 
the region of the mixed phase in the core of hybrid stars.
}

\FullConference{29th Johns Hopkins Workshop on current problems in particle theory: strong matter in the heavens\\
		 1-3 August\\
		 Budapest}

\begin{document}

\section{Introduction}

It is widely believed that quark matter exists in
high-temperature and/or high-density situations like in the relativistic
heavy-ion collisions (RHIC) \cite{rhic1} or in the core of neutron stars \cite{mad3,chen}, 
and the ``deconfinement transition'' has been actively searched.
Theoretical studies using model calculations or based on the first
principle, lattice QCD \cite{rev}
have been also carried out by many authors to find the critical temperature. 
Although many exciting results have been reported, the deconfinement transition
has not been clearly understood yet. 
 
We, hereafter, consider the phase transition from hadron phase to
three-flavor quark phase in neutron-star matter in a
semi-phenomenological way. 
Since many theoretical calculations have suggested that the deconfinement 
transition should be of first order
in low-temperature and high-density area \cite{pisa,latt},  
we assume the first-order phase transition in this paper. 
If the deconfinement transition is of first order, one may expect a 
{\it mixed phase} during the transition. Actually, 
the hadron-quark mixed phase has been considered during the
hadronization era in RHIC \cite{rhic21,rhic22,rhic23}
or in the core region of neutron
stars \cite{gle2,web1,web2}. 

There is an issue about the mixed phase for the first-order phase
transitions with two or more chemical potentials \cite{gle1}.
We often use the Maxwell construction (MC) to derive the equation
of state (EOS) in thermodynamic equilibrium, as in the liquid-vapor phase transition
of water, where both phases consist of single particle species.
However, if many particle species participate in the phase
transition as in neutron-star matter, MC is no more an appropriate method. 
Before Glendenning first pointed out \cite{gle1}, 
many people have applied MC to get EOS for the  
first-order phase transitions \cite{weis,migd,elli,rose} expected in
neutron stars, such
as pion or kaon condensation and the deconfinement transition. 

\begin{figure}[h!]\begin{center}
\includegraphics[width=80mm]{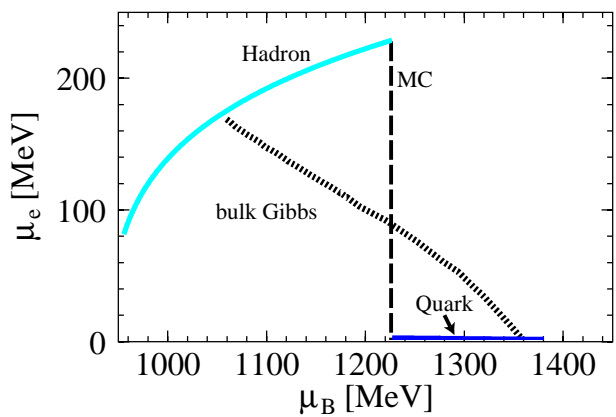}
\caption{ Phase diagram in the $\mu_\mathrm{B}-\mu_\mathrm{e}$ plane. 
There appears no region of the mixed phase by the calculation with MC, while a wide region of the mixed
		   phase by the ``bulk Gibbs'' calculation.}
\label{bulk}
\end{center}
\end{figure}

Let us consider a deconfinement transition in neutron-star matter. 
We must introduce many chemical potentials for particle species, 
but the independent ones in this case are reduced to two,
i.e.\ baryon-number chemical potential $\mu_\mathrm{B}$ and 
charge chemical potential $\mu_\mathrm{Q}$,
due to chemical equilibrium and total charge neutrality.
They are nothing but the neutron and electron chemical potentials, $\mu_n$ and
$\mu_e$, respectively. 
In the mixed-phase these chemical potentials should be spatially constant. 
When we naively apply MC to get EOS in thermodynamic equilibrium, we
immediately notice that $\mu_\mathrm{B}$ is constant in the mixed-phase, while
$\mu_\mathrm{e}$ is different in each phase because of the difference of 
the electron number density in these phases. 
This is because MC uses EOS of bulk matter in each phase,
which is of locally
charge-neutral and uniform matter; many electrons are needed in hadron
matter to compensate the positive charge of protons,
while in quark matter charge
neutrality is almost fulfilled without electrons.
Thus
\begin{equation}
\mu_{\mathrm{B}}^{\mathrm{Q}} = \mu_{\mathrm{B}}^{\mathrm{H}}, \hspace{10pt}
\mu_{\mathrm{e}}^{\mathrm{Q}} \neq  \mu_{\mathrm{e}}^{\mathrm{H}},
\label{chemeqMC}
\end{equation}
in MC, where superscripts ``Q'' and ``H'' denote the quark and hadron phases, respectively.
Glendenning emphasized that we must use the Gibbs conditions (GC)
in this case instead of MC, which relaxes the charge-neutrality
condition to be globally satisfied as a whole, not locally in each phase \cite{gle1}. 
GC imposes the following conditions,
\begin{eqnarray}
 \mu_{\mathrm{B}}^{\mathrm{Q}} &= \mu_{\mathrm{B}}^{\mathrm{H}},
 \hspace{5pt} \mu_{\mathrm{e}}^{\mathrm{Q}} =
 \mu_{\mathrm{e}}^{\mathrm{H}}, \nonumber \\
P^{\mathrm{Q}} &=P^{\mathrm{H}} , \hspace{5pt} T^{\mathrm{Q}} = T^{\mathrm{H}}.
\end{eqnarray}
He demonstrated a wide region of the mixed phase, where two phases
have a net charge but totally charge-neutral:
EOS thus obtained,
different from
that given by MC,
never exhibits a constant-pressure region.
He simply considered
a mixed phase consisting of two bulk matters separated by a sharp
boundary without any surface
tension and the Coulomb interaction, which we call ``bulk Gibbs'' for convenience.
``Bulk Gibbs'' requires that each matter can have a net charge but the total charge
is neutral, 
\begin{equation}
 f_V \rho_{\mathrm{ch}}^\mathrm{Q} + (1-f_V) \rho_{\mathrm{ch}}^\mathrm{H} = 0,
\end{equation}
where $f_V$ means the volume fraction of quark matter in the mixed phase
and ``$\rho_\mathrm{ch}^\mathrm{Q,H}$'' means charge density in each matter.
Figure \ref{bulk} shows the phase diagram in the $\mu_\mathrm{B}$ - $\mu_e$ plane.
We can see that there is a discontinuous jump in $\mu_e$ for the case of MC, 
while the curve given by ``bulk Gibbs'' smoothly connects uniform hadron matter
and uniform quark matter; the mixed phase can appear in a wide
$\mu_\mathrm{B}$ region in ``bulk Gibbs'', in contrast with MC \cite{alf2}.

 However, this ``bulk Gibbs'' is too simple to study the mixed phase,
 since we must consider matter with non-uniform structures
instead of two bulk uniform matters; the mixed phase should have
 various geometrical structures where both the number and charge
 densities are no more uniform. 
Then we have to take into account 
 finite-size effects like the surface and the Coulomb interaction energies.
We show that the mixed phase should be narrow in
 the $\mu_\mathrm{B}$ space by the charge
 screening effect, and derive EOS for the deconfinement transition in
 neutron-star matter. 
 We shall see EOS results in being similar to that given by MC. 
 We also discuss the
 interplay of the Coulomb interaction effect and the surface effect in the context
 of the hadron-quark mixed phase.
Preliminary results for the droplet case has been already
reported in Ref.\ \cite{end1}.

\section{Brief review of the previous works}\label{history}

\begin{wrapfigure}{l}{35mm}
\begin{center}
\includegraphics[width=35mm,keepaspectratio]{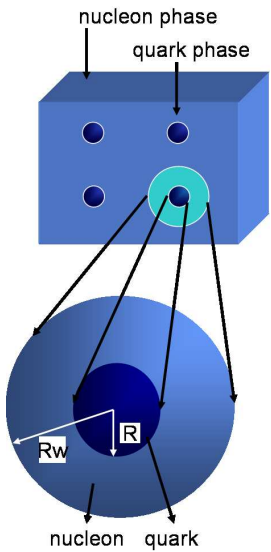}
\caption{Wigner-Seitz approximation for the droplet case. $R$ is the
 droplet radius and $R_\mathrm{W}$ the cell radius.}
\label{wsapp}
\end{center}
\end{wrapfigure}

 Heiselberg et al.\ \cite{pet} studied a geometrical structure in the mixed
 phase by considering the spherical quark droplets embedded in hadron matter.
 They introduced the surface tension and treated its strength as a free
 parameter  because the
 surface tension at the hadron-quark interface has not been clearly understood. 
 They pointed out that if the surface tension parameter $\sigma$ is large
 ($\sigma \geq$ 90 MeV/fm$^2$), the region of the mixed phase is largely
 limited  or cannot exist. 
\begin{figure}[h!]\begin{center}
\includegraphics[width=.6\textwidth]{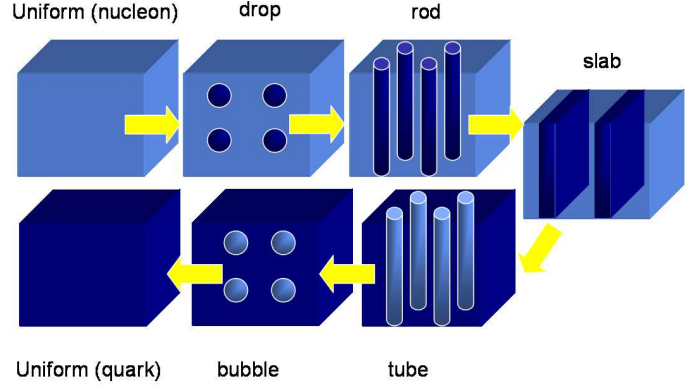}
\caption{Schematic image of structured mixed phase.}
\label{smp}
\end{center}
\end{figure}
Subsequently 
Glendenning and Pei \cite{gle2} have suggested ``crystalline structures of the mixed phase'' 
which have some geometrical structures, ``droplet'', ``rod'', ``slab'', ``tube'',
and ``bubble'' (Fig.\ \ref{smp}),
assuming a small $\sigma$\cite{gle2,gle3}.
The finite-size effects are obvious in these calculations by observing energies.
We may consider only a single cell, by dividing the whole space into 
equivalent Wigner-Seitz cells: the cell size is denoted by $R_W$ and the
size of the lump (droplet, rod, slab, tube or 
bubble) by $R$ (Fig.\ \ref{wsapp}). 
Then the surface energy density 
is expressed in terms of the surface tension parameter $\sigma$ as 
\begin{eqnarray}
\epsilon_\mathrm{S} = \frac{f_V \sigma d}{R},
\end{eqnarray}
where $d$ denotes the
dimensionality of each geometrical structure; 
$d=3$ for droplet and bubble,
$d=2$ for rod and tube, and $d=1$ for slab. 
The Coulomb energy density reads
\begin{eqnarray}
\epsilon_\mathrm{C} &=& 2\pi e^2 \left(\rho_\mathrm{ch}^\mathrm{H}
- \rho_\mathrm{ch}^\mathrm{Q}  \right)^2 R^2 \Phi_d(f_V) \hspace{5pt},\\
\Phi_d(f_V) &\equiv& \left[2 (d-2)^{-1} \left( 1 -\frac{1}{2} d f_V^{1-2/d}  \right)
+ f_V  \right] \left(d+2 \right)^{-1},
\end{eqnarray}
where we simply assume an uniform density distribution in each phase.
When we minimize the sum of $\epsilon_S$ and $\epsilon_C$ with respect
to the size $R$ for a given volume fraction $f_V$, we can get the 
well-known relation, 
\begin{equation}
\epsilon_S = 2 \epsilon_C.
\label{edensCS}
\end{equation}
This implies that an
optimal size of the lump is determined by the balance of these finite-size effects.
Eventually we can express
the sum of the surface and Coulomb energy densities 
with an optimal cell size \cite{pet}:
\begin{equation} 
\epsilon^{(d)}_\mathrm{C} + \epsilon^{(d)}_\mathrm{S} = 3 f_V d \left(
							      \frac{\pi
							      \sigma^2
							      \left(
							       \rho_\mathrm{ch}^\mathrm{H} - \rho_\mathrm{ch}^\mathrm{Q}  \right)^2 \Phi_d (f_V)}{2d}  \right)^{1/3}. 
\label{bulk_cs} 
\end{equation}
Thus we can calculate the energy of any geometrically structured mixed phase with
(\ref{bulk_cs}) by changing the parameter $d$. 
Many authors have taken this treatment for the mixed
phase\cite{gle2,alf2,pet}.
Note that the energy sum in Eq.\ (\ref{bulk_cs}) becomes larger as the
surface tension gets stronger, while the relation  Eq.(\ref{edensCS}) is always kept. 
%

%

However, this treatment is not a self-consistent but a perturbative one,
since the charge screening effect for the Coulomb potential or the
rearrangement effect of
charged-particles in the presence of the Coulomb interaction is completely discarded. 
We shall see that the Coulomb potential is never weak in the mixed
phase, and thereby this treatment overestimates the Coulomb energy.
The charge screening effect is included only if we introduce 
the Coulomb potential and consistently
solve the Poisson equation with other equations of motion for charged particles. 
Consequently it is a highly non-perturbative effect.
Norsen and Reddy \cite{nors} have studied the Debye screening effect in
the context of kaon condensation to see a large change of the
charged-particle densities like kaons and protons. 
Maruyama et al.\ have numerically studied it in the context of liquid-gas phase
transition at subnuclear densities \cite{maru1}, where nuclear pasta can be
regarded as geometrical structures in the mixed phase. 
Subsequently, we have also studied kaon
condensation at high-densities \cite{maru2}, where 
kaonic pasta structures appeared in the mixed phase. 
Through these works we have figured out the role of
the Debye screening in the mixed phase. 

Voskresensky et al.\ \cite{vos} explicitly  studied the Debye screening
 effect for a few geometrical structures of the hadron-quark mixed phase.
 They have shown that
 the optimal value of the size of the structure cannot be obtained
 due to the charge screening, if the surface tension is not so small.
 They called it mechanical instability.
 It occurs because the Coulomb energy density is suppressed for  
 the size larger than the Debye screening length (cf.\ Eq.~(\ref{lambda})).
They also suggested that the properties of the mixed phase become very
 similar to those given by MC, if the charge screening effectively works.
The apparent violation of the Gibbs conditions in MC (Eq.~(\ref{chemeqMC}))
 can be remedied by including the Coulomb potential in a
 gauge-invariant way: the number density of charged particles is given by a
 gauge-invariant combination of the chemical potential and the Coulomb
 potential, and thereby it 
 can be different in each phase for a
 constant charge chemical potential if the 
 the Coulomb potential takes different values in both phases.
However, they used a linear approximation to solve the Poisson equation analytically.

If the Coulomb interaction effect is so large, it would be important
to study it without recourse to any approximation.
In this paper we numerically study the charge screening effect on the
structured mixed phase during the deconfinement transition in neutron-star matter  
in a self-consistent way.
Actually we shall see importance of non-linear effects included in the
Poisson equation. 

\section{Self-consistent calculation with the Coulomb potential}\label{formalism}

We have presented our formulation in detail in Ref.\ \cite{end2,end3}. Here
we only briefly explain it.
 We consider the geometrically structured mixed phase (SMP) where one
 phase is embedded in the other phase with a certain geometrical shape.
We divide the whole space into equivalent charge-neutral Wigner-Seitz
cells.  
The Wigner-Seitz cell is a representative of equally divided space
under approximation of a geometrical symmetry: sphere in three
dimensional (3D) case, rod in 2D case and slab in 1D case 
with the size $R_W$. The cell consists of 
two phases in equilibrium: one phase with a size $R$ is embedded in the
other phase 
as illustrated in Fig.\ \ref{wsapp}.

Quark phase consists of {\it u}, {\it d}, {\it s} quarks and
electron.  We incorporate the MIT Bag model and assume the sharp boundary at the
hadron-quark interface: 
{\it u} and {\it d}
quarks are treated to be massless and {\it s} is massive
($m_s=150$MeV), and they interact with each other by the
one-gluon-exchange interaction inside the bag.
Hadron phase consists of proton, neutron and electron. The
effective potential is used to describe the interaction
between nucleons and to reproduce the saturation properties of nuclear matter.

Then total thermodynamic potential ($\Omega_\mathrm{total}$)  
consists of the hadron, quark and electron contributions and the
surface contribution:
\begin{equation}
\Omega_\mathrm{total} = \Omega_\mathrm{H} +\Omega_\mathrm{Q}
 +\Omega_\mathrm{S}, 
\label{ometot}
\end{equation}
where $\Omega_\mathrm{H(Q)}$ denotes the contribution of hadron(quark)
phase. We here introduce the surface contribution $\Omega_\mathrm{S}$,
parametrized by the 
surface tension parameter $\sigma$, $\Omega_\mathrm{S} = \sigma S$ with 
$S$ being the area of the interface.
Note that it may be closely related with the confining mechanism and unfortunately 
we have no definite idea about how to incorporate it.  
Actually many authors
have treated its strength as a free parameter and seen 
how the results are changed by its value\cite{gle2,alf2,pet}, which we
also follow in this report.

To write down the thermodynamic potential we use the idea of the density functional
 theory (DFT) within the local density approximation (LDA)
 \cite{parr,drez}. 
The Coulomb interaction energy is included in $\Omega_\mathrm{H(Q)}$ and can be
 expressed in terms of
particle densities,
\begin{equation}
E_V = \frac{1}{2} \sum_{i,j} \int_{V_\mathrm{W}} d^3 r d^3 r^{\prime} \frac{Q_i \rho_i(\vec{r}) Q_j \rho_j(\vec{r}^{\prime})}{\left| \vec{r} - \vec{r}^{\prime} \right|},
\end{equation}
where $i=u,d,s,p,n,e$ with $Q_i$ being the particle charge ($Q=-e < 0$
for the electron) and $V_\mathrm{W}$ the volume of Wigner-Seitz cell.
Accordingly the Coulomb potential is defined as
\begin{equation}
V (\vec{r}) = -\sum_i \int_{V_\mathrm{W}} d^3 r^{\prime} \frac{e Q_i
 \rho_i(\vec{r}^{\prime})}{\left| \vec{r} - \vec{r}^{\prime} \right|}+V_0,
\label{vcoul}
\end{equation}
where $V_0$ is an arbitrary constant representing the gauge degree of
 freedom. 
We fix the gauge by a condition $V(R_W) = 0$ in this paper. 
Operating a Laplacian $\nabla^2$ on the Coulomb potential
$V(\vec{r})$, we automatically derive the Poisson equation.

There are introduced six chemical potentials in Eq.~(\ref{ometot}) 
corresponding to particle species in the cell.
Then we have to determine now eight variables, i.e., 
six chemical potentials, $\mu_\mathrm{i} (i=u,d,s,p,n,e)$, 
and  the radii $R$ and $R_W$.
Fixing $R$ and $R_W$, 
we have four conditions due to chemical equilibrium in each phase and at
the interface:
\begin{eqnarray}
&& \mu_\mathrm{u}+\mu_\mathrm{e} = \mu_\mathrm{d},\nonumber\\
&& \mu_\mathrm{d} = \mu_\mathrm{s}, \nonumber \\
&& \mu_\mathrm{p}+\mu_\mathrm{e} = \mu_\mathrm{n} \equiv \mu_\mathrm{B}, \nonumber \\
&& \mu_\mathrm{n} = \mu_\mathrm{u}+2 \mu_\mathrm{d}, \nonumber\\
&& \left(\mu_\mathrm{p} = 2 \mu_\mathrm{u}+\mu_\mathrm{d}\right).
\label{chemeq}
\end{eqnarray}
Therefore, once two chemical potentials $\mu_\mathrm{B}$ and
$\mu_\mathrm{e}$ are given, 
we can determine other four chemical potentials, 
$\mu_\mathrm{u}$, $\mu_\mathrm{d}$, $\mu_\mathrm{s}$ and $\mu_\mathrm{p}$. 
We determine $\mu_\mathrm{e}$ by the global charge neutrality
condition: 
\begin{equation}
 f_V \rho_{\mathrm{ch}}^{\mathrm{Q}} + (1-f_V) \rho_{\mathrm{ch}}^{\mathrm{H}} = 0,
\end{equation}
where the volume fraction $f_V=\left(\frac{R}{R_W}\right)^d$, and $d$
denotes the dimensionality of each geometrical structure.
At this point $f_V$ is still fixed.
The particle densities $\rho_i$ are given by the equations of motion,
which are derived form the extremum 
condition, $\frac{\delta\Omega_\mathrm{tot}}{\delta\rho_i}=0$, 
\begin{equation}
\mu_i = \frac{\delta E_\mathrm{kin+str}}{\delta \rho_i (\vec{r})} - N_i V(\vec{r}),
\label{mu_i}
\end{equation}
where $E_\mathrm{kin+str}$ stands for the kinetic and strong interaction
energy except the Coulomb interaction energy.

Then we find the optimal value of $R$ ($R_W$ is fixed and 
thereby $f_V$ is changed by $R$) 
by using one of GC;
\begin{equation}
P^\mathrm{Q} = P^\mathrm{H} + P_\sigma, 
\label{pbalance}
\end{equation}
where the pressure coming from the surface tension is given by 
\begin{equation}
P_{\sigma}= \sigma \frac{d S}{d V_{\rm Q}}.
\end{equation}
The pressure in each phase $P^\mathrm{Q(H)}$ 
is given by the thermodynamic relation: 
$P^\mathrm{Q(H)}=-\Omega_\mathrm{Q(H)}/V_\mathrm{Q(H)}$,
where $\Omega_\mathrm{Q(H)}$ is the thermodynamic potential in each
phase and given by adding electron and the Coulomb interaction
contributions to $\Omega_\mathrm{Q(H)}$ in Eqs.\ (\ref{ometot}).
Finally, we determine $R_W$ by minimizing thermodynamic potential. 
Thus, once
$\mu_\mathrm{B}$ is given, all other values $\mu_i$ ($i=u,d,s,p,e$)
and $R$, $R_W$ can be obtained. Accordingly, the particle density
profiles are given by Eq.~(\ref{mu_i}).

We numerically solve Eqs.~(\ref{mu_i}) and the Poisson equation in a self-consistent way.
In numerical calculation, every point inside the cell is represented by a grid point
(the number of grid points $N_\mathrm{grid} \approx 100 $). 
Equations of motion are solved by
a relaxation method for a given baryon-number chemical potential under
constraints of the global charge neutrality. 
Note that we keep GC throughout the numerical procedure.

\section{Numerical results and discussions}\label{results}

\begin{figure}[h!]\begin{center}
\includegraphics[width=80mm]{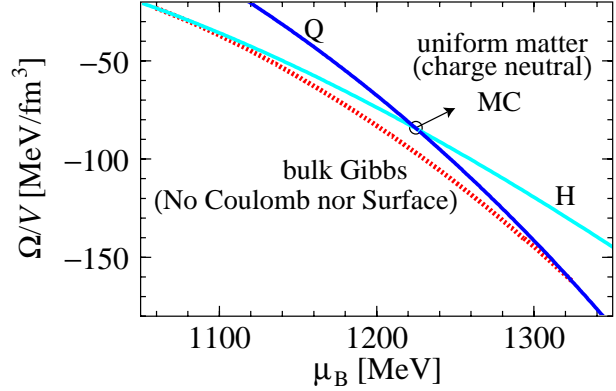}
\caption{ Thermodynamic potential density ($\Omega/V = \omega$) in the $\mu_\mathrm{B}$ space.}
\label{mubome00}
\end{center}
\end{figure}

First, we present the thermodynamic potential density of each uniform
matter and that given by 
``bulk Gibbs'' in Fig.\ \ref{mubome00}. 
In uniform matter, hadron phase is thermodynamically favorable for 
$\mu_\mathrm{B} < 1225$ MeV and quark phase above it.
Therefore
we plot $\delta \omega$, difference of the thermodynamic potential density
between the mixed phase and each uniform matter:
\begin{equation}
\delta \omega = \begin{cases} \omega_\mathrm{total}-\omega^\mathrm{uniform}_\mathrm{H} \quad \mu_\mathrm{B}
\leq 1225 \mathrm{MeV}, \\
\omega_\mathrm{total}-\omega^\mathrm{uniform}_\mathrm{Q} \quad \mu_\mathrm{B} \geq 1225 \mathrm{MeV},
\end{cases}
\end{equation}
where $\omega_\mathrm{total}=\Omega_\mathrm{total}/V_W$, etc.

\begin{figure}[htb]
\begin{minipage}[t]{70mm}
\includegraphics[width=70mm]{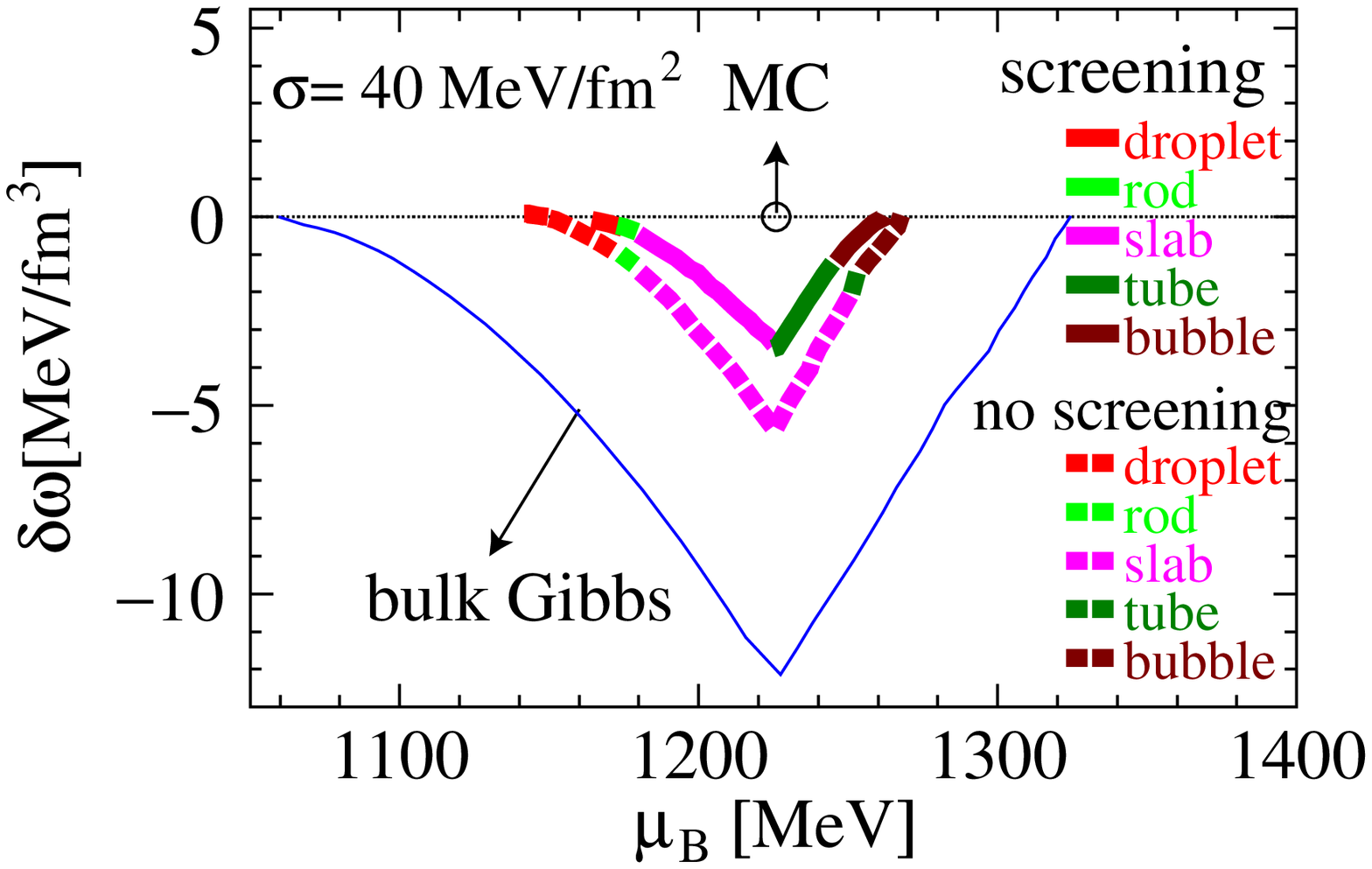}
\caption{Difference of the thermodynamic potential density
 as a function of baryon-number chemical potential $\mu_\mathrm{B}$ for
 $\sigma=40$ MeV/fm$^2$. If $\delta \omega$ is negative, the mixed phase
 is a thermodynamically favorable state. MC determines one point of phase transition in uniform matter,
denoted as a circle in the $\mu_B$-$\delta\omega$ plane.}
\label{omed40}
\end{minipage}
\hspace{8pt}
\begin{minipage}[t]{70mm}
\includegraphics[width=70mm]{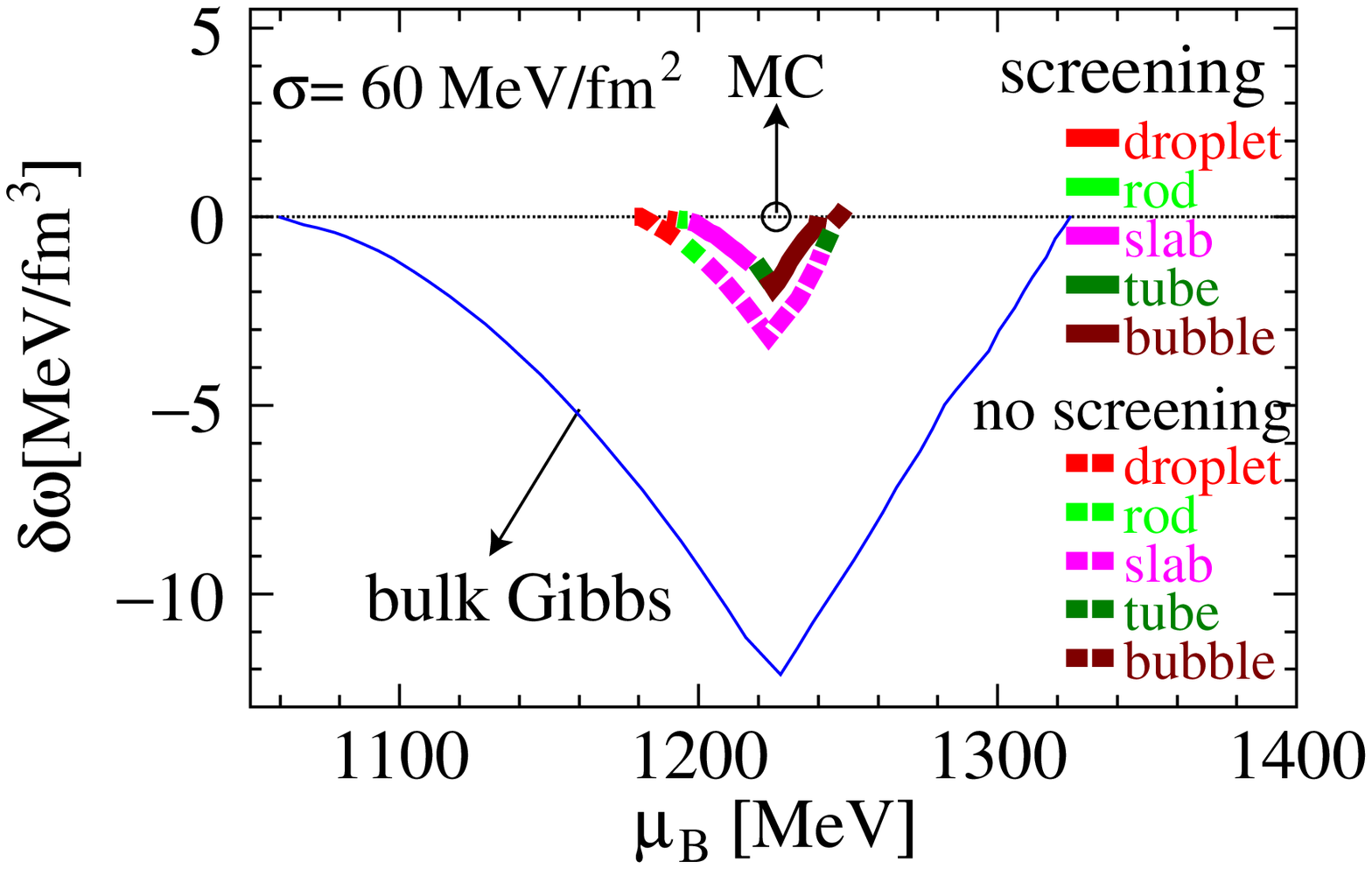}
\caption{Same as Fig.\ \protect\ref{omed40} for $\sigma=60$ MeV/fm$^{2}$. The negative $\delta
 \omega$ region is narrower than the $\sigma=40$ MeV/fm$^{2}$ case.}
\label{omed60}
\end{minipage}
\end{figure}

The results are shown in Figs.\ \ref{omed40} and \ref{omed60}.
In these figures we also depict two results for comparison: 
one is given by``bulk Gibbs'', 
where the finite-size effects are completely discarded. 
The other is the thermodynamic potential given by a perturbative
treatment of the Coulomb interaction, which is denoted by ``no screening''; 
discarding the Coulomb potential $V(\vec{r})$, 
we solve the equations of motion to get {\it uniform} density
profiles (no rearrangement), 
then evaluate the Coulomb interaction energy by
using the density profiles thus determined. Note that ``no screening''
violates the gauge invariance, while our treatment respects it \cite{vos}.
Since chemical potential itself is gauge variant, we have to include
$V(\vec{r})$ together like in Eq.\ (\ref{mu_i}) to satisfy the gauge
invariance: when the Coulomb potential is
shifted by a constant value, $V(\vec{r}) \Longrightarrow V({\vec{r}}) - V_0$,
the charge chemical potential should be also shifted as $\mu_i \Longrightarrow
\mu_i+N_i V_0$. Incidentally, the phase diagram in the $\mu_\mathrm{B}-\mu_\mathrm{e}$ plane
(see, e.g.,
Fig.~\ref{bulk}) is not well-defined by this reason, while many people
have written it \cite{alf2,gle5}.
We can see the screening effects by comparing 
the results given by ``no screening'' with those by the self-consistent calculation denoted by ``screening''.
$\delta \omega$ given by MC appears as a point denoted by a circle in
Figs.\ \ref{omed40} and \ref{omed60} where two conditions, $P^\mathrm{Q}=P^\mathrm{H}$ and
$\mu_\mathrm{B}^\mathrm{Q}=\mu_\mathrm{B}^\mathrm{H}$, are satisfied.
On the other hand the mixed phase derived from ``bulk Gibbs'' appears in a
wide region of $\mu_{\mathrm{B}}$. 
Therefore, if the region of the mixed phase becomes
narrower, it signals that the properties of the mixed phase become close
to those given by MC. 
One may clearly see that $\omega_{\mathrm{total}}$ becomes close to that
given by MC due to the finite-size effects, the effects of the surface
tension and the Coulomb interaction.

The large increase of $\delta\omega$ from the ``bulk Gibbs'' curve comes from the effects
of the surface tension and the Coulomb potential. Since the surface
tension parameter is introduced by hand, we must carefully study the
effects of the surface tension and the Coulomb interaction, separately. 
 From the difference between the result
given by ``no screening'' and that by ``bulk Gibbs'', we 
can roughly say that about $2/3$ of the increase comes from the
effect of the surface tension and $1/3$ from the Coulomb interaction
(see Eq. (\ref{edensCS})). 
Comparing the result of self-consistent calculation with that of ``no screening'',
we can see that the change of energy 
caused by the screening effect is not so large
, but still the same order of
magnitude as that given by the surface effect. 

%
\begin{figure}[htb]
\begin{minipage}[t]{70mm}
\includegraphics[width=70mm]{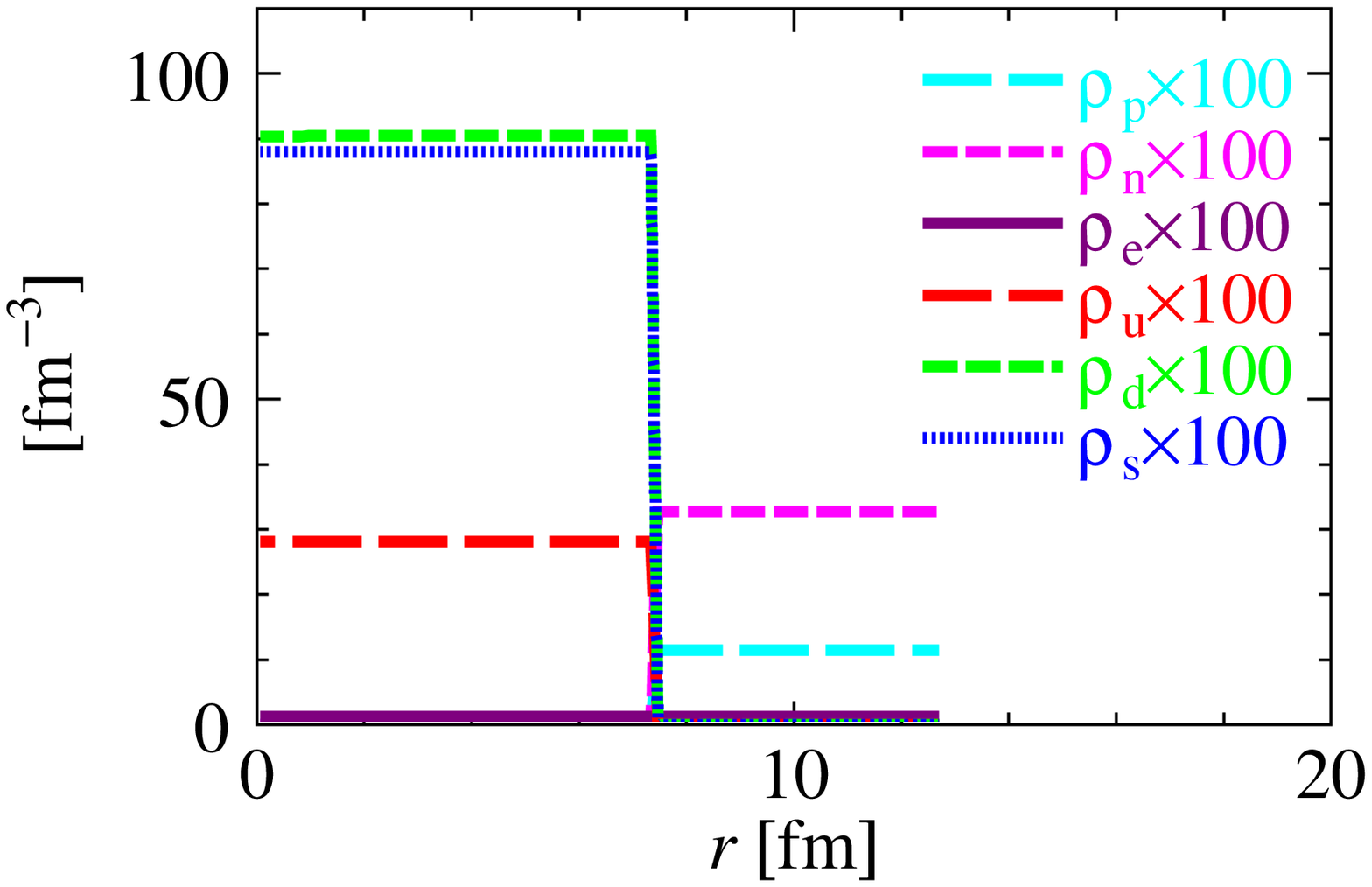}
\caption{Density profiles in the droplet phase given by ``no screening'' 
for $\mu_B=1189$ MeV and $\sigma=60$ MeV/fm$^{2}$. 
 They are uniform in each phase. 
$R=7.2$ fm and $R_W=12.8$ fm.}
\label{densprof-no}
\end{minipage}
\hspace{8pt}
\begin{minipage}[t]{70mm}
%
\includegraphics[width=70mm]{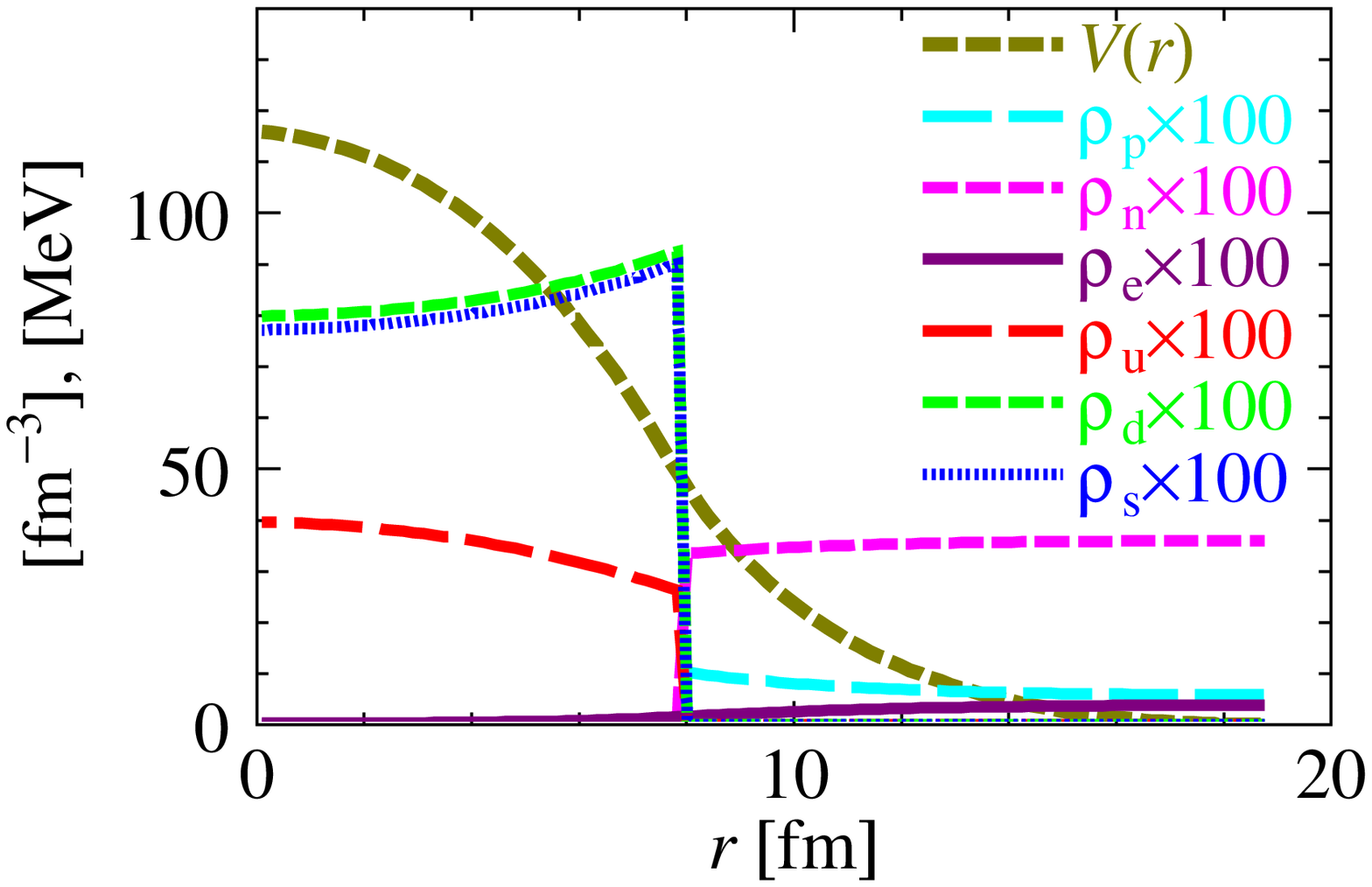}
\caption{Density profiles and the Coulomb potential given by the self-consistent
 calculation for the same parameter set as Fig.~5. $R=7.7$ fm and
 $R_W=18.9$ fm.}
\label{densprof-sc}
\end{minipage}
\end{figure}
%
If the surface tension is stronger, the relative importance of the
charge screening effect becomes smaller and the effect of the surface tension
becomes more prominent, as is seen in Figs.\ \ref{omed40} and \ref{omed60}.

\begin{wrapfigure}{r}{70mm}
\includegraphics[width=70mm]{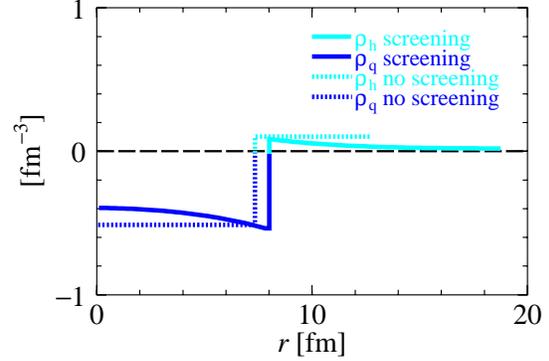}
\caption{Local-charge densities for the ``no screening'' case and the
 case of the self-consistent with the screening effect. For ``no screening'', charge
 density of each phase is constant over the region. The absolute
 value of the charge density is
 larger than that given by the self-consistent calculation in each phase. 
 In the hadron phase, the charge density becomes almost 
 vanished near the cell boundary $r=R_W$.}
\label{chdensprof}
\end{wrapfigure}

\begin{figure}[htb]
\begin{minipage}[t]{70mm}
\includegraphics[width=70mm]{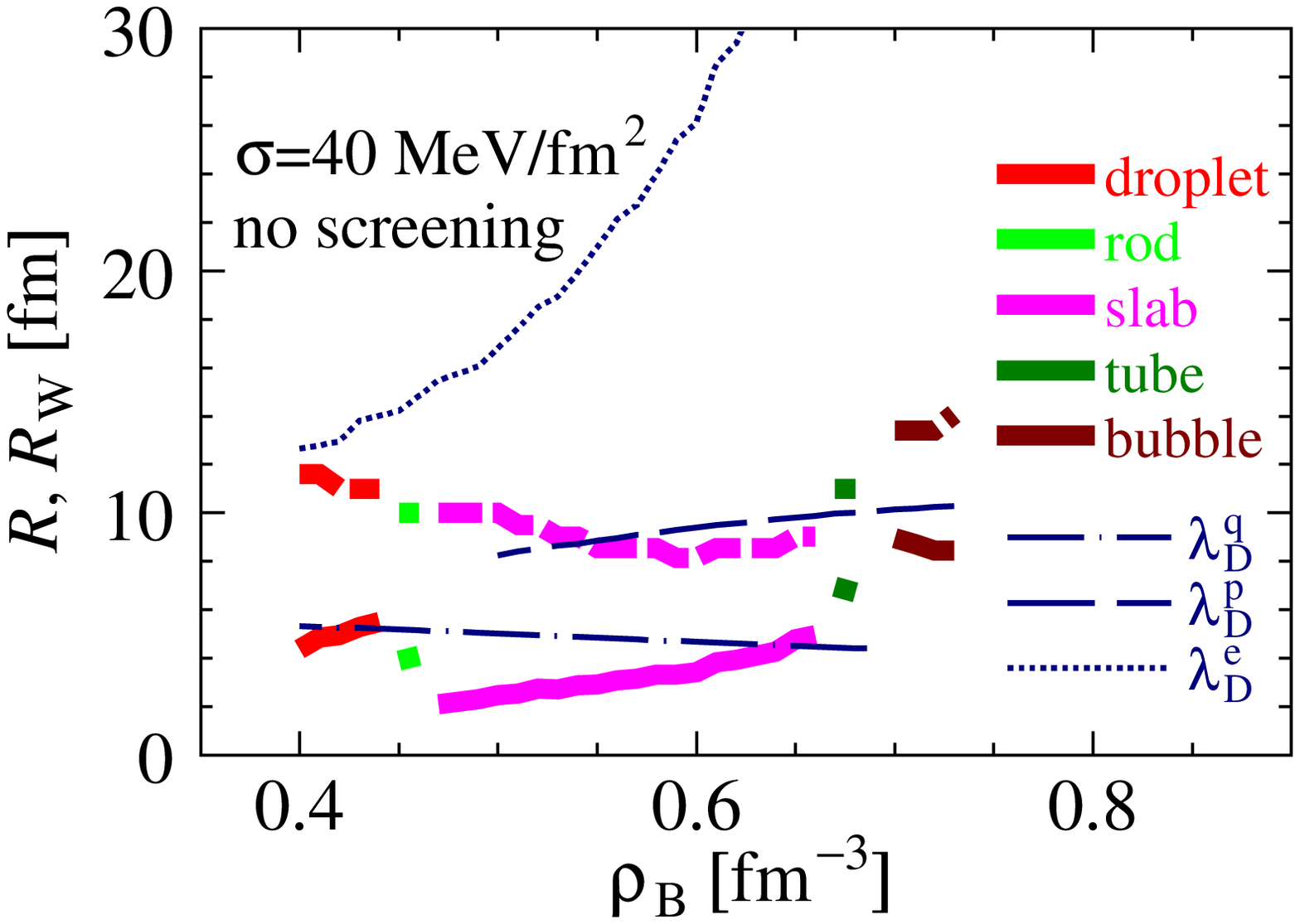}
\caption{Lump and cell radii given by the ``no screening'' 
 calculation. The Debye screening length is also depicted for
 comparison. $R$ is thick-solid line and $R_W$ is thick-dashed line. 
We can see the size of the structure becomes less than the
 Debye screening length.}
\label{rho-cell40no}
\end{minipage}
\hspace{8pt}
\begin{minipage}[t]{70mm}
\includegraphics[width=70mm]{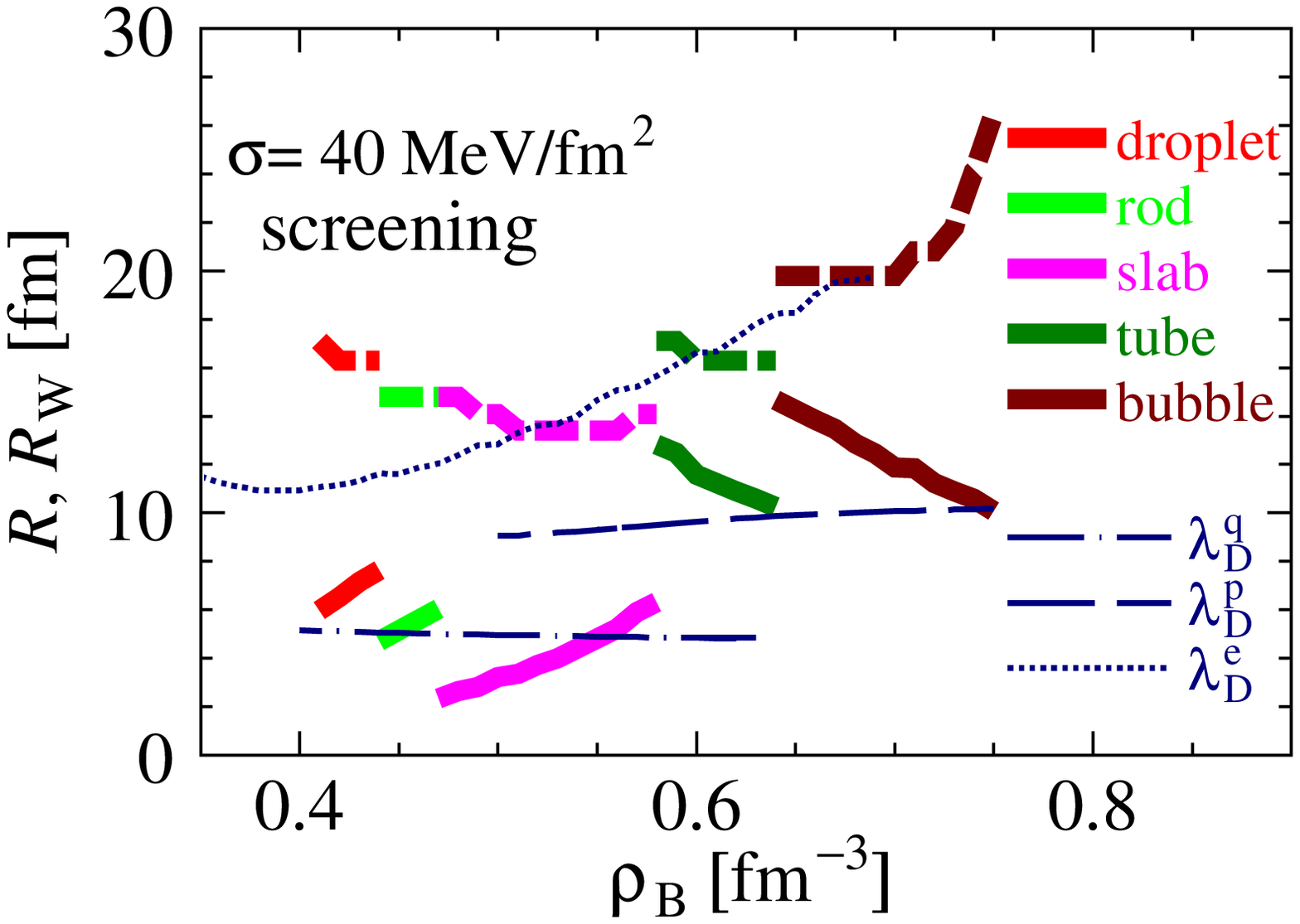}
\caption{Same as Fig.\ \protect\ref{rho-cell40no} given by the self-consistent calculation with the
 screening effect. The size of the structure becomes larger than 
that given by ``no screening'', and consequently exceeds the 
 Debye screening length.}
\label{rho-cell40sc}
\end{minipage}
\end{figure}

Although the charge screening has not so large effects on bulk properties of the matter, 
we shall see that it is remarkable for the charged particles to change the properties of the mixed phase. 
The screening effect induces
the rearrangement of the charged particles. 
We can see this effect by comparing 
Fig.\ \ref{densprof-no} with Fig.\ \ref{densprof-sc}.
The quark
phase is negatively charged and the hadron phase is positively charged. 
The negatively charged particles in the quark phase such as {\it d},
{\it s}, {\it e}  and the positively charged particle in the hadron phase {\it p} are
attracted toward the boundary.
On the contrary the positively charged particle in the quark phase {\it u}
and  negatively charged particle in the hadron phase {\it e} are repelled
from the boundary.
The charge screening effect also reduces the net charge in each phase.
In Fig.\ \ref{chdensprof}, we show the local charge densities
of the two cases shown in Figs.\ \ref{densprof-no} and \ref{densprof-sc}.
The change of the number of charged particles due to the screening is as follows:
In the quark phase, the numbers of $d$ and $s$ quarks and
electrons decrease,
while the number of $u$ quark increases. 
In the hadron phase, on the other hand, 
the proton number should decrease and the electron number
should increase. 
Consequently the local charge decreases in the both phases.
In Fig.\ \ref{chdensprof} we can see that the core region of the droplet
tends to be charge-neutral and 
 near the boundary of the Wigner-Seitz cell 
is almost charge-neutral.
In Figs.\ \ref{rho-cell40no} and \ref{rho-cell40sc} we present the lump
and cell radii for each density.
As we have shown in the previous paper\cite{vos}, the Coulomb energy
is suppressed for larger $R$ by the screening effect. 
The $R$ dependence of the total thermodynamic potential comes from the
contributions of the surface tension and the Coulomb interaction: 
the optimal radius giving the minimum of the thermodynamic potential 
is then determined by the balance between two
contributions, since the
former gives a decreasing function, while the latter an increasing one.
If the Coulomb energy is
suppressed, the minimum of the thermodynamic potential is shifted
to larger radius. 
As a result the size of the embedded phase ($R$) and the cell size ($R_W$) become large.
In Ref.~\cite{vos} they demonstrated that the minimum disappears for
a large value of the surface tension parameter: the structure becomes 
mechanically unstable in this case. 

%
\begin{figure}
\begin{center}
\includegraphics[width=120mm]{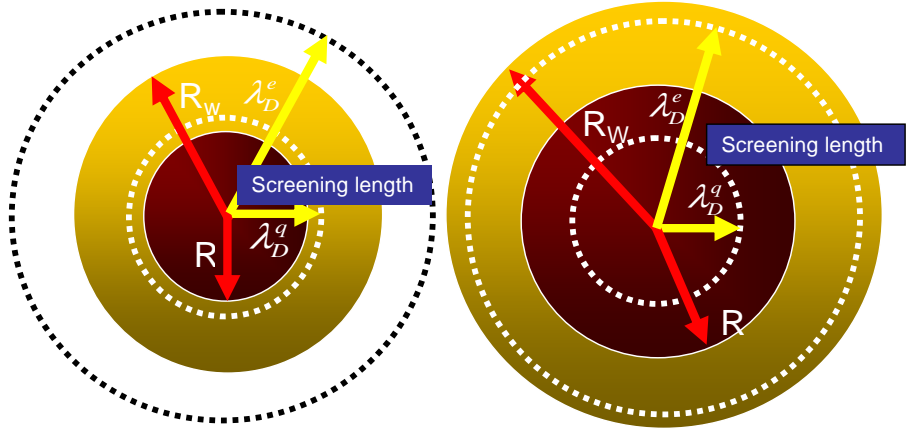}
\caption{Schematic graphs of the droplet size and the Debye screening
 length. Right figure shows the case of the self-consistent calculation with the
 screening effect and left figure ``no screening''.}
\label{lambdadrop}
\end{center}
\end{figure}
%
%
We cannot show it directly in our framework because such unstable solutions 
are automatically excluded during the numerical procedure, 
while we can see its tendency 
in Figs.\ \ref{rho-cell40no} and \ref{rho-cell40sc}:
$R$ and $R_W$ get larger by the screening effect.

We also see the relation between the size of the geometrical structure and the Debye
screening length.
The Debye screening length appears in the {\it linearized} Poisson
equation 
and is then given as 
\begin{equation}
\left(\lambda^{q}_D\right)^{-2}\!\!=\! 4 \pi \sum_f Q_f \! \left( \frac{\partial \langle
				     \rho_f^\mathrm{ch}
				     \rangle}{\partial \mu_f} \right), \hspace{5pt}
\left(\lambda^{p}_D\right)^{-2}\!\!=\! 4 \pi Q_p \! \left( \frac{\partial \langle
				     \rho_p^\mathrm{ch}
				     \rangle}{\partial \mu_p} \right), \hspace{5pt}
\left(\lambda^{e}_D\right)^{-2}\!\!=\! 4 \pi Q_e \! \left( \frac{\partial \langle
				     \rho_e^\mathrm{ch}
				     \rangle}{\partial \mu_e} \right), 
\label{lambda}
\end{equation}
where $\langle \rho_f^\mathrm{ch} \rangle$ stands for 
the averaged density in quark phase, $\langle \rho_p^\mathrm{ch}
\rangle$ is proton number averaged density in the hadron phase 
and $\langle \rho_e^\mathrm{ch} \rangle$
is the electron charge density averaged inside the cell.
It gives a rough measure for the screening effect:
At a distance larger than the Debye screening length, 
the Coulomb interaction is effectively suppressed. 

In Fig.\ \ref{rho-cell40no} we show sizes of geometrical structure
for ``no screening'' case.
If we ignore the screening effect, the size of the embedded phase is
comparable or smaller than the corresponding quark Debye screening
length $\lambda_D^q$ (Fig.\ \ref{lambdadrop}). 
This may mean that the Debye
screening is not so important.
Actually, many authors have neglected the screening
effect due to this argument\cite{alf2,pet}. 
In Fig.\ \ref{rho-cell40sc}, however, 
we see that the size of the embedded phase 
can be larger than $\lambda_D^q$ (Fig.\ \ref{lambdadrop})
in the self-consistent calculation.
We can also see the similar situation about $R_W$ and $\lambda_D^e$.
This means that the screening has important
effects in this mixed phase. 
We cannot expect such a effect
without solving the Poisson equation because of the non-linearity.

\begin{figure}[htb]
\begin{minipage}[t]{70mm}
\includegraphics[width=70mm]{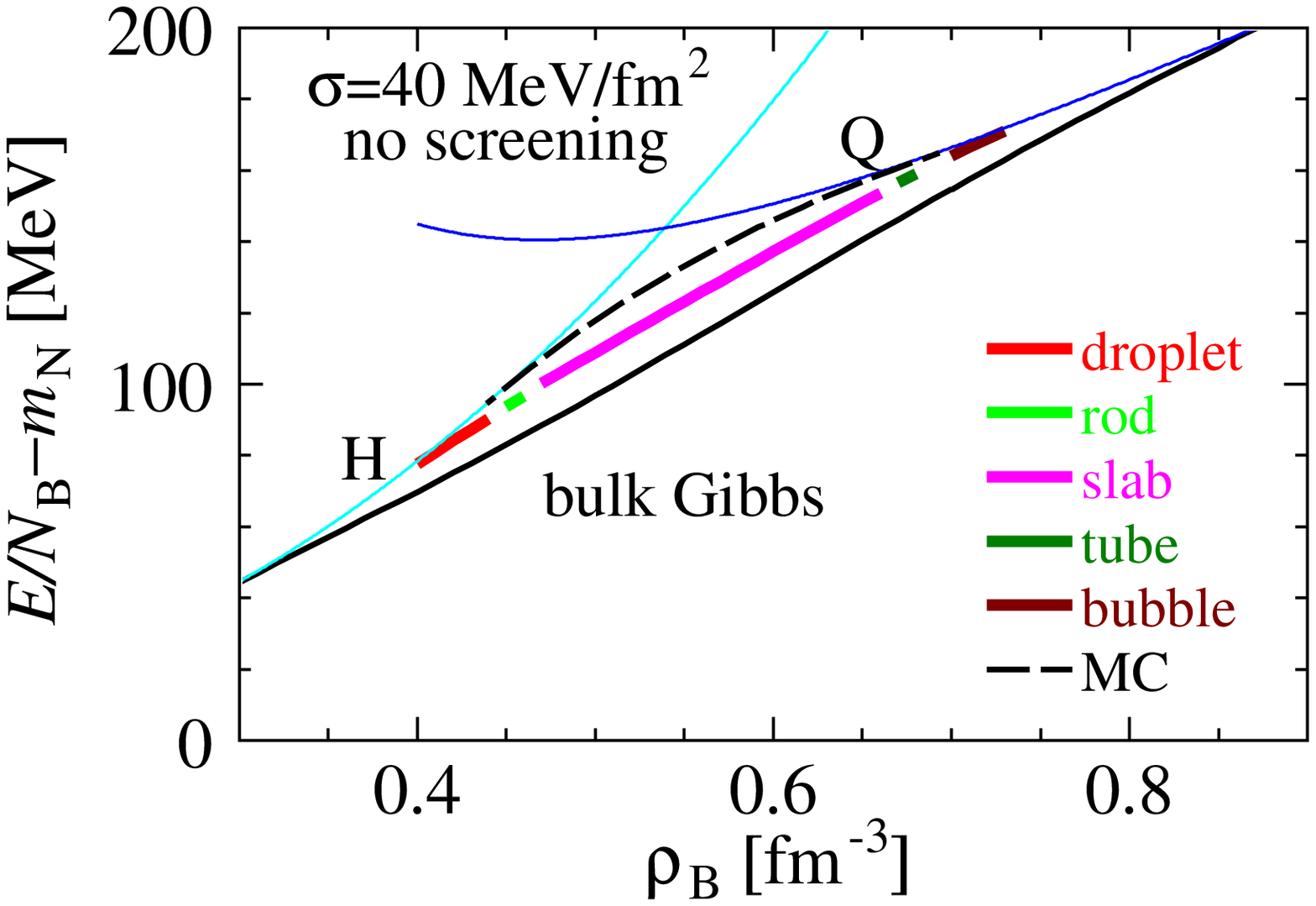}
\caption{Equation of state for $\sigma=40$ MeV/fm$^2$
 without screening effect.}
\label{eos40sc}
\end{minipage}
\hspace{8pt}
\begin{minipage}[t]{70mm}
\includegraphics[width=70mm]{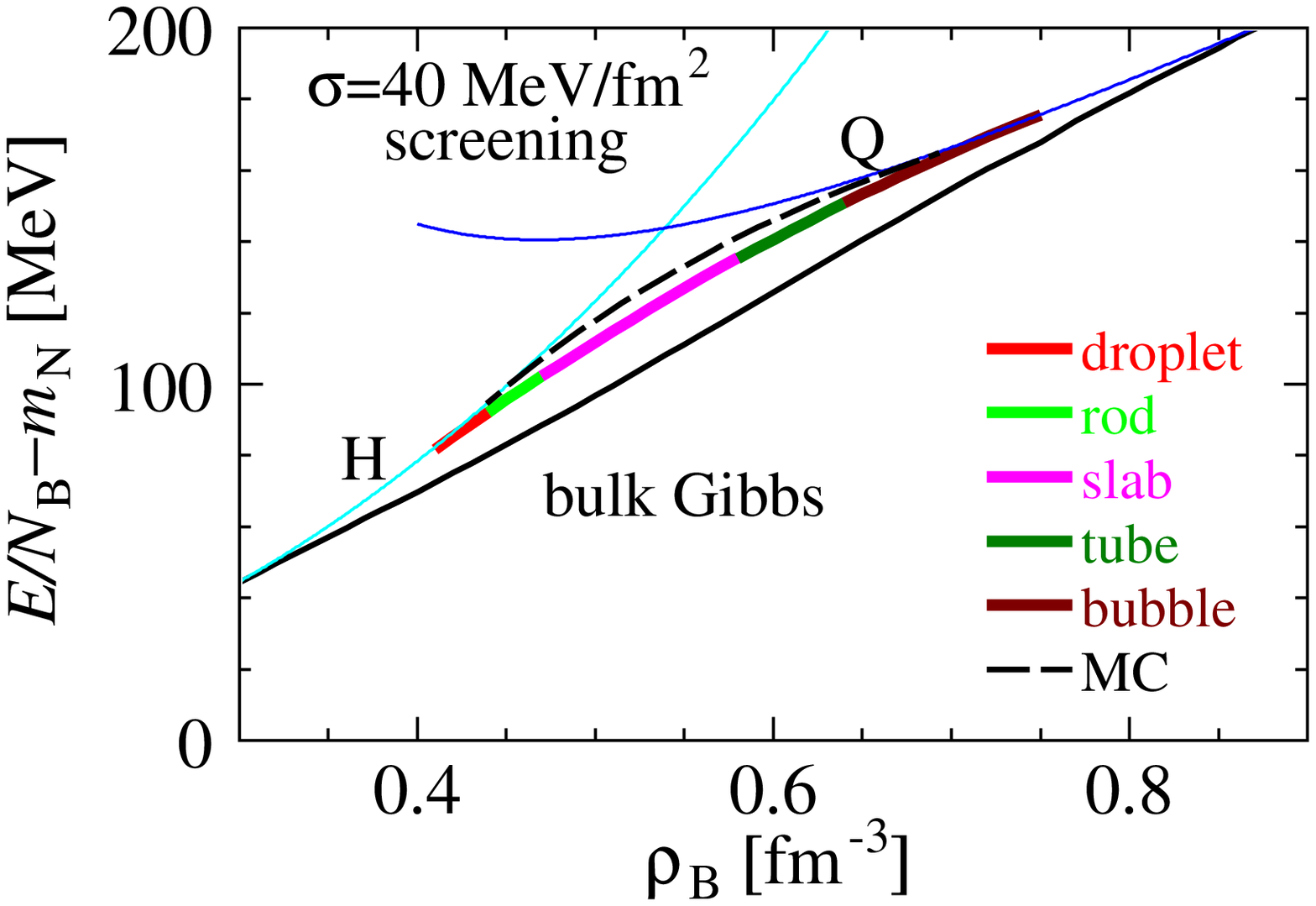}
\caption{Same as Fig.\ \protect\ref{eos40sc} for ``screening''.}
\label{eos40no}
\end{minipage}
\end{figure}

\begin{figure}[htb]
\begin{minipage}[t]{70mm}
\includegraphics[width=70mm]{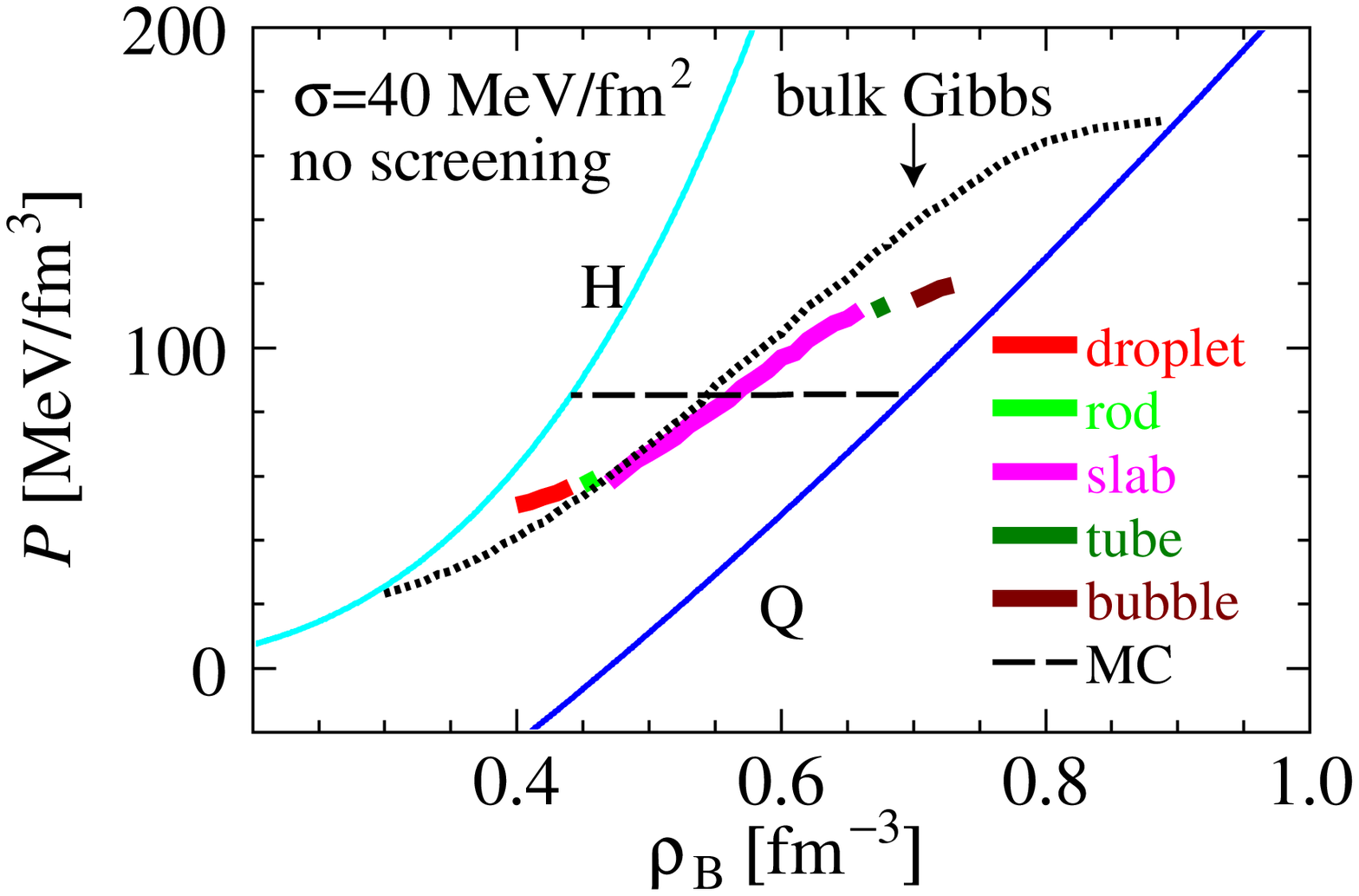}
\caption{Pressure as a function of baryon-number density given by the
 ``no screening'' calculation for $\sigma=40$MeV/fm$^2$. The results given
 by ``bulk Gibbs'' and MC are also presented for comparison. }
\label{pres40no}
\end{minipage}
\hspace{8pt}
\begin{minipage}[t]{70mm}
\includegraphics[width=70mm]{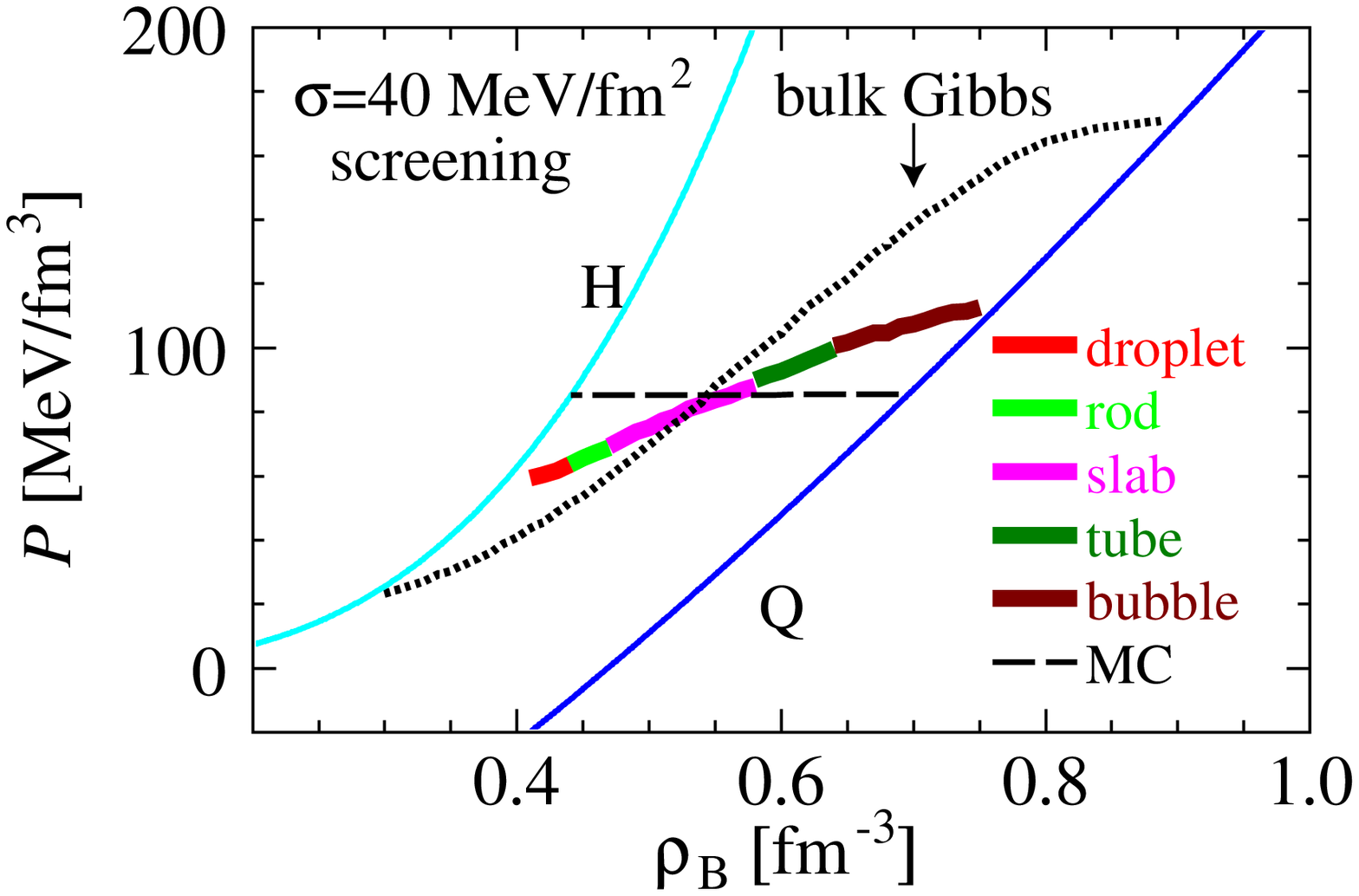}
\caption{Same as Fig.\ \protect\ref{pres40no} given by the self-consistent calculation with the screening effect.}
\label{pres40sc}
\end{minipage}
\end{figure}

We present EOS in Figs.\ \ref{eos40sc} and \ref{eos40no}.
We can see that EOS of two cases become close to that given by MC
by including the finite size effects. Moreover, EOS of ``screening''
becomes more similar to that given by MC than that of ``no screening''.
We show the pressure results in Figs.\ \ref{pres40no} and
\ref{pres40sc}, which is given by 
$P=\rho_\mathrm{B}^2\frac{\partial
(E/N_\mathrm{B})}{\partial \rho_\mathrm{B}}$ or  $ P=-\Omega/V$: the
former is given by using Figs.~\ref{eos40sc} and \ref{eos40no} and the
latter by using the results of Fig.\ \ref{omed40}.
The pressure becomes more similar to that
given by MC by the charge screening effect, 
which shows a larger pressure near the beginning 
and weaker one near the termination of the phase transition.

We have used 
the surface tension parameter $\sigma$ in the present study.
Surface tension is a very difficult problem because it should be
self-consistent with the two phases of matter, quark and hadron.
Lattice QCD, based on the first principle, 
predicts that $\sigma$ may be
10-100 MeV/fm$^2$\cite{kaja,huan}.
Although this result is for high temperature case,
our choice is  within it.
Moreover, the results given by other model calculations
\cite{mad1,mad2,berg,mond} 
are similar to our value.

Let us briefly consider some implication of these our results for compact star phenomena.
Glendenning\cite{gle2} suggested many SMP appear
in the core region by using ``bulk Gibbs'':
the mixed phase should appear for several kilometers. 
However we can say that the region of SMP
should be narrow in the $\mu_\mathrm{B}$ space and EOS is more similar
to that of MC 
due to the finite-size effects.
These results look to be consistent with those given by other studies.
Bejger et al.\ \cite{bejg} have
examined the relation between the mixed phase
and glitch phenomena, and shown that the mixed phase should be narrow if the
glitch is generated by 
the mixed phase in the inner core. 
On the other hand the gravitational
wave asks for density discontinuity in the core region\cite{mini}.

\section{Summary and concluding remarks}\label{summary}

We have studied
the charge screening effect in 
the hadron-quark mixed phase 
in neutron-star matter.
We have elucidated the charge screening effect 
comparing the results of ``no screening'' 
with those of ``screening''  which includes non-linearity of the Poisson equation.
The density profiles of the charged
particles are drastically modified by the screening effect, while the
thermodynamic potential is not affected so much; the charge
rearrangement induced by the screening effect tends to make the net
charge smaller in each phase. Consequently the system tends to be
locally charge-neutral, which suggests 
that the Maxwell construction (MC) is 
effectively justified even if it is thermodynamically incorrect.
In this context, it would be interesting to refer to the work by 
Heiselberg \cite{hei}, who  studied the screening effect on a quark
droplet (strangelet) in the vacuum,
and suggested the importance of the rearrangement of charged particles.

We have seen that thermodynamic quantities such as thermodynamic potential and
energy become close to those derived from MC by the screening effect,
which also suggests that MC is effectively justified due to the
screening effect. 
As another case of the system with two or more chemical potentials, 
kaon condensation  has been also studied \cite{maru1} and the results are shown 
to be similar to those in the present study. 
Thus the importance of the screening effect should be
a common feature for the first-order phase transitions in high-density matter.

We have included the surface tension at the hadron-quark interface, 
while its definite value is not clear at present. 
An uncertainty in the value of the surface tension does not allow to
conclude whether the mixed phase exists or not. 
There are many estimations for the surface tension at the hadron-quark
interface in lattice QCD \cite{kaja,huan}, in shell-model
calculations \cite{mad1,mad2,berg} and in model
calculations based on the dual-Ginzburg Landau theory \cite{mond}.
Our parameter is in that reasonable range.

We have considered some implications of our results for neutron star phenomena.
The screening effect would restrict
the allowed region of the structured mixed phase in neutron stars, 
in contrast with a wide region
given by ``bulk Gibbs''\cite {gle2,gle3}. 
It could be said that they should change the bulk 
property of neutron stars, especially the structure of the core region. 

Compact stars
have the strong magnetic field and its origin is not well understood. One
possibility is that it comes from ferromagnetism of the quark matter in the
core \cite{tat1,tat2,tat3}. Therefore it should be interesting to 
include the magnetic field contribution in our formalism.
We have assumed zero temperature here. 
It would be  much interesting to include the finite-temperature
effect. Then it is possible to draw the phase diagram in the
$\mu_\mathrm{B}$ - $T$ plane and we can study the properties of the
deconfinement phase transition; our study may be extended to treat 
the mixed phase to appear during the hadronization of QGP in the
nucleus-nucleus collisions 
and supernova explosions.

 In this study we have used a simple model for quark matter to
figure out the finite-size effects in SMP. However,
it has been suggested that the color superconductivity would be a ground state of quark
matter \cite{alf2,alf1,alf3}. Hence we will include it in a further study.
The hadron phase should be also treated more realistically; for example, 
we should include the
hyperons or kaons in hadron matter.
In the recent studies the
mixed phase has been also studied \cite{neum,shov,redd} in the context of
various phases in the color superconducting phase.

\section*{Acknowledgements}
We wish to acknowledge Dr.\ T.~Tanigawa for fruitful discussion. 
T.~E.  would like to acknowledge Dr.\ M.~Alford and Dr.\ F.~Weber 
for their useful comments and discussions during this conference. 
This work is supported by the Grant-in-Aid for the 21st Century
COE ``Center for the Diversity and Universality in Physics'' at Kyoto
University from the Ministry of Education, Culture, Sports, 
Science and Technology (MEXT) of Japan.

\end{document}